\documentclass[a4paper,10pt]{article}
\usepackage[T1]{fontenc}
\usepackage[margin=1in]{geometry}
\usepackage{graphicx,relsize,libertine,afterpage,amsmath,amssymb,textcomp,xcolor,soul,sfmath,authblk,mathtools,multicol,float,gensymb,enumitem,gensymb}

\usepackage{bm}

\definecolor{dgreen}{rgb}{0,0.6,0} \definecolor{dred}{rgb}{0.6,0,0}
\definecolor{dpurple}{HTML}{A020F0} \definecolor{dblue}{rgb}{0,0,1}
\definecolor{hlcolor}{rgb}{1,1,0.8}
\definecolor{change}{rgb}{0,0.6,0}

\usepackage{lineno}

\usepackage[colorlinks=true,linkcolor=dblue,citecolor=dblue,urlcolor=dblue]{hyperref}
\usepackage[super,sort&compress,comma]{natbib}

\makeatletter
\renewcommand{\maketitle}{\bgroup\setlength{\parindent}{0pt}
\begin{flushleft}
  \bgroup\sffamily\huge\textbf{\@title}\egroup
  
  \bgroup\sffamily\textbf{\@author}\egroup
\end{flushleft}\egroup
}
\makeatother

\usepackage{xr}
\makeatletter
\newcommand*{\addFileDependency}[1]{% argument=file name and extension
  \typeout{(#1)}
  \@addtofilelist{#1}
  \IfFileExists{#1}{}{\typeout{No file #1.}}
}
\makeatother

\def\headline#1{{\hrulefill\quad\lower.3em\hbox{#1}\quad\hrulefill}}

 \sethlcolor{hlcolor}

\usepackage{float}
\newfloat{efigure}{htp}{efig}
\floatname{efigure}{Extended Data Fig.}

\usepackage[font={sf,footnotesize}]{caption}
\DeclareCaptionLabelSeparator{vert}{ |}
\usepackage[nameinlink]{cleveref}
\Crefname{equation}{Eq.}{Eqs.}
\Crefname{figure}{Fig.}{Figs.}
\Crefname{efigure}{Extended Data Fig.}{Extended Data Figs.}
\crefname{efigure}{Extended Data Fig.}{Extended Data Figs.}
\Crefname{tabular}{Tab.}{Tabs.}
\crefrangelabelformat{equation}{#3#1#4-#5#2#6}
\creflabelformat{equation}{#2#1#3}

\usepackage[explicit]{titlesec}

\titleformat{name=\section,numberless}[block]{\large\bfseries\sffamily}{}{0em}{#1}[]

\titleformat{name=\subsection,numberless}[block]{\normalsize\bfseries\sffamily}{}{0em}{#1}[]

\titlespacing{name=\section ,numberless}{0pt}{10pt plus 2pt minus 2pt}{0pt plus 2pt minus 2pt}
\titlespacing{name=\subsection ,numberless}{0pt}{5pt plus 2pt minus 2pt}{0pt}

\providecommand{\dd}{\text{d}}

\providecommand{\fun}[2]{#1\!\left(#2\right)}

\providecommand{\cgiven}{\vert}

\providecommand{\normal}[1]{\fun{\mathcal{N}}{#1}}

\newcommand{\OO}{\boldsymbol{0}}

\newcommand{\II}{\mathbf{I}}

\newcommand{\WW}{\mathbf{W}}

\newcommand{\xx}{\mathbf{x}}

\newcommand{\yy}{\mathbf{y}}
\renewcommand{\dd}{\mathsf{d}}
\renewcommand{\AA}{\mathbf{A}}

\newcommand{\CC}{\mathbf{C}}

\newcommand{\sx}{\sigma^2_\mathrm{x}}

\newcommand\e{\text{E}}

\renewcommand\i{\text{I}}

\newcommand{\RN}[1]{%
  \textup{\uppercase\expandafter{\romannumeral#1}}%
}

\DeclarePairedDelimiter\floor{\lfloor}{\rfloor}

\providecommand{\abs}[1]{\left|#1\right|}

\newcommand{\methods}{\hyperref[sec:online_methods]{Methods}}
\newcommand{\discussion}{\hyperref[sec:discussion]{Discussion}}

\newcommand{\inputnoise}{process noise}

\title{Bridging physiological and perceptual views of autism by means of sampling-based Bayesian inference}
\author[1@]{Rodrigo Echeveste}
\author[1*]{Enzo Ferrante}
\author[1*]{Diego H. Milone}
\author[2*]{In\'es Samengo} 
\affil[1]{\small Research Institute for Signals, Systems and Computational Intelligence sinc(i) (FICH-UNL/CONICET), Santa Fe, Argentina.}
\affil[2]{\small Medical Physics Department and Balseiro Institute (CNEA-UNCUYO/CONICET), Bariloche, Argentina. }
\affil[*]{{\small equal contribution,} $^@${\small corresponding author: recheveste@sinc.unl.edu.ar}}

\begin{document}
\maketitle

% comment out for word count
%\linenumbers
% uncomment for word count
% \hyphenpenalty=10000
% \exhyphenpenalty=10000

% \tableofcontents
%Abstract

\sffamily
\noindent
% Up to 150 words
\rightline{\textit{Accepted for publication in Network Neuroscience}}\\

\textbf{Theories for autism spectrum disorder (ASD) have been formulated at different levels: ranging from physiological observations to perceptual and behavioral descriptions. Understanding the physiological underpinnings of perceptual traits in ASD remains a significant challenge in the field. Here we show how a recurrent neural circuit model which was optimized to perform sampling-based inference and displays characteristic features of cortical dynamics can help bridge this gap. The model was able to establish a mechanistic link between two descriptive levels for ASD: a physiological level, in terms of inhibitory dysfunction, neural variability and oscillations, and a perceptual level, in terms of hypopriors in Bayesian computations. We took two parallel paths: inducing hypopriors in the probabilistic model, and an inhibitory dysfunction in the network model, which lead to consistent results in terms of the represented posteriors, providing support for the view that both descriptions might constitute two sides of the same coin.}

\label{sec:introduction}

Autism spectrum disorder (ASD) refers to a complex neurodevelopmental
condition involving persistent challenges in social interaction and communicative skills, and restricted/repetitive behaviors \cite{american2013diagnostic}. While some recent studies suggest that ASD could be detected during the first year of life in some children, early signs seem to be non-specific, with group differences more robustly found after children's first birthday (see Ref.~\citenum{ozonoff2008onset} for a review). 

Almost two decades ago, John Rubenstein and Michael Merzenich suggested that many of the symptoms related to ASD might reflect an abnormal ratio between excitation and inhibition leading to hyper-excitability of cortical circuits in ASD subjects\cite{rubenstein2003model}. Since then, a variety of studies have linked reduced inhibitory signaling in the brain with ASD symptoms, either observing how behavior typically associated with ASD emerges in animals when inhibitory pathways are altered, or measuring gamma-aminobutyric acid (GABA) concentration or GABA receptors in several brain regions (see Ref.~\citenum{cellot2014gabaergic} for a detailed review). Further support for this view comes from the fact that ASD patients suffer from epilepsy with a prevalence up to 25 times that of the neurotypical population\cite{bolton2011epilepsy}. 

Establishing a direct link between ASD and impaired inhibition in specific circuits in humans has not been easy. Indeed, two recent in-vivo studies in humans have shown puzzling results\cite{robertson2016reduced,horder2018gabaa}. In these studies inhibition was assessed both behaviorally (in visual tasks where inhibition is widely believed to play a key role in neurotypical behavior) and by measuring either GABA concentration\cite{robertson2016reduced} or number of GABA receptors\cite{horder2018gabaa} in the brains of ASD and control subjects. Interestingly, while ASD subjects showed a marked deficit in binocular rivalry, characteristic of a disruption in inhibitory signaling, GABA concentrations in the visual cortex were normal\cite{robertson2016reduced}. However, while GABA concentration was predictive of rivalry dynamics in controls, the same was not true within the ASD population, evidencing a disruption of inhibitory action. Similarly, while ASD subjects show an altered performance in the paradoxical motion perception task (a proxy measure of GABA signaling), GABA receptor availability in the brain of those participants showed no significant difference from controls\cite{horder2018gabaa}. Both studies suggest an impairment in inhibitory signaling which cannot be explained by coarse differences in GABA concentration or receptor availability at the level of brain areas, and which might affect specific circuits instead. To complicate matters further, there is evidence for not only inhibitory but also excitatory disfunction in ASD, and it has been hypothesized that homeostatic principles might be the reason behind this seemingly contradictory result\cite{nelson2015excitatory}. The idea being that if, for instance inhibition is reduced, excitatory synapses might be then adjusted to try to compensate for the overall change in neural activity that reduction would ensue. Computational modeling of local cortical circuits expressed in terms of excitation and inhibition might therefore provide a fruitful avenue of research to guide future experiments. 

From the point of view of perception in ASD, a variety of theories have been put forward over the last two decades. Highly influential descriptive theories include: the weak central coherence theory\cite{happe2006weak} and the enhanced perceptual functioning theory\cite{mottron2006enhanced}. Here we will focus on computational accounts of perception in ASD, and in particular on a Bayesian view of perception\cite{palmer2017bayesian}. We will later also make connections to another influential computational theory formulated in terms of predictive coding\cite{van2013predictive,van2014precise}.

Within the Bayesian framework, inference about the external world proceeds by multiplicatively combining pre-existent knowledge (expressed in terms of a \emph{prior} probability distribution) and current sensory evidence (represented in terms of a \emph{likelihood} function), to form a \emph{posterior} distribution which encapsulates our belief about the state of the world after having observed a given stimulus\cite{knill1996perception}. Rather than expressing that belief as a single point estimate of what is most probable, the posterior distribution provides a richer description, naturally incorporating the associated uncertainty which remains after the observation. A growing body of evidence indicates that, at least in some settings, the brain is able to operate with probability distributions in this way to perform approximate Bayesian inference (see Ref.~\citenum{fiser2010statistically}, for a review). In recent years it has been proposed that in ASD subjects these forms of Bayesian computations are carried out abnormally: overweighting sensory evidence with respect to prior information\cite{pellicano2012world,palmer2017bayesian}. Concretely, the authors in Ref.~\citenum{pellicano2012world} proposed that this is a consequence of chronically attenuated priors (termed \emph{hypopriors}), characterized by broader distributions (i.e. higher uncertainty). 

The related theoretical framework of predictive coding proposes that the cortex is organized following a circuit motif where feedback connections from higher- to lower-order sensory areas signal predictions of lower-level responses, while feedforward connections signal errors between predictions and actually observed lower-level responses\cite{rao1999predictive}. Proponents of predictive coding theories have rightfully pointed out that Bayesian theories by themselves (without specifying a concrete implementation) do not offer a mechanistic explanation for ASD perception\cite{van2013predictive}, which is key to understand how physiological observations may be linked to perceptual and behavioral traits in ASD subjects. As has been observed by Ref.~\citenum{aitchison2017or}, Bayesian inference and predictive coding are not necessarily mutually exclusive: predictive coding can be seen as a computational motif which can implement several computational goals (one of which is Bayesian inference), while Bayesian inference can be seen as a computational objective which can have several implementations (one of which is predictive coding). Moreover, as noted in the aforementioned review, telling apart the use of a Bayesian predictive coding scheme from a direct variable code in an empirical setting is no trivial matter. Strong transient overshoots at stimulus onset, for instance, which are a typical signature of predictive coding, can also emerge in direct variable coding schemes\cite{aitchison2016hamiltonian,echeveste2020cortical}. Indeed, while weighting predictive errors more strongly by increasing synaptic gains in the motif could explain sensory hypersensitivity in ASD subjects\cite{palmer2017bayesian}, a competing explanation can be provided within a direct variable coding scheme, as we show in the present study. We note however that while predictive coding schemes can incorporate gamma oscillations\cite{bastos2012canonical}, it is not clear how they would account for the contrast-dependent frequency modulation of these oscillations\cite{roberts2013robust}, or the stimulus-dependent modulations of neural variability\cite{churchland2010stimulus,orban2016neural}.

\begin{figure}[h] 
\centering\includegraphics[width=0.4\textwidth]{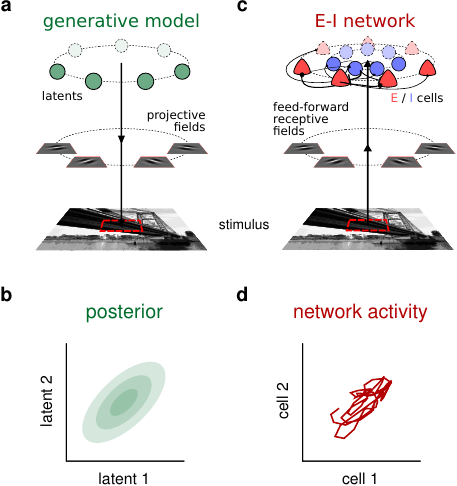}
\caption{\label{fig_sketch}
\textbf{Sketches of the generative model, and a neural circuit implementing sampling-based probabilistic inference under that model.} \textbf{a,}~The Gaussian scale mixture (GSM) generative model. Under this model, each image patch is built as a linear combination of local features (projective fields), whose intensities are drawn from a multivariate Gaussian distribution. This linear combination is then further scaled by a global contrast level and subject to noise. The features were in this case a set of localized oriented Gabor filters which differed only in their orientations and were uniformly spread between $-90^\circ$ and $90^\circ$. The image serving as stimulus in the figure is for illustration only. Photo Credit: Santa Fe Bridge by Enzo Ferrante \url{https://eferrante.github.io/}) 
\textbf{b,}~2D projection of the posterior distribution for a given a visual stimulus as computed by the Bayesian ideal observer under the GSM.
\textbf{c,}~The recurrent E--I neural network receives an image patch as an input, which is filtered by feedforward receptive fields matching the projective fields of GSM in \textbf{a}.  Each latent variable in the GSM is represented by the activity of one E cell in the network. \textbf{d,}~2D projection of the neural responses of E cells corresponding the same 2 latent variables shown in \textbf{b}. Over time, the network samples from posterior distribution corresponding to the stimulus it receives.
}
\end{figure}

A popular implementation choice for probabilistic inference is that of probabilistic population codes (PPCs)\cite{ma2006bayesian}, where the posterior distribution is encoded in the average rates of a population of neurons. This framework has been used in the past to link inhibitory deficits and Bayesian computations in an artificial neural network model consisting of two feed-forward layers followed by a stage of divisive normalization\cite{rosenberg2015computational}. In this work, a probabilistic version of the model was constructed to capture the ``oblique effect''. This term describes the fact that neurotypical subjects tend to be more sensitive to cardinal than to oblique orientations in a visual orientation discrimination task\cite{westheimer1998orientation}. Indeed, a modulation of the divisive normalization factor in this model was shown to account for the observed reduction of the oblique effect in ASD subjects\cite{dickinson2014oblique}. The standard PPC framework requires constant Fano factors (no variability modulation)\cite{ma2006bayesian}, and furthermore feed-forward network implementations can only capture mean rate responses, but fail to account for the dynamical properties of neural responses that arise from recurrent connectivity. It is hence unclear in this framework how altered neural variability observed in the ASD population\cite{milne2011increased,haigh2015cortical} and gamma oscillations\cite{van2015increased} would relate to probabilistic computations in these subjects.

Sampling-based theories for probabilistic inference offer an alternative mechanistic implementation for Bayesian inference. Within this framework, neural circuits represent posterior distributions by drawing samples over time from those distributions\cite{berkes2011spontaneous,haefner2016perceptual}. Interestingly, sampling-based models for probabilistic inference have recently begun to establish direct links between cortical dynamics and perception (see Ref.~\citenum{echeveste2020cortical}). A neural circuit model of a cortical hypercolumn respecting Dale´s principle and performing fast sampling-based inference in a visual task displayed a suite of features which are typically observed in cortical recordings across species and experimental conditions. The network showed highly variable responses with strong inhibition-dominated transients at stimulus onset, and stimulus-dependent gamma oscillations, as observed in the cortex\cite{haider2013inhibition,ray2010differences,roberts2013robust}. The model further evidenced stimulus-dependent variability modulations consistent with experimental findings\cite{roberts2013robust}. Divisive normalization of mean responses\cite{carandini2012normalization} was also shown to emerge in this network as a result of its recurrent dynamics. This is interesting since divisive normalization was precisely the starting point for the probabilistic model in Ref.~\citenum{rosenberg2015computational}, and in previous work linking uncertainty and neural variability via gain modulation\cite{henaff2020representation}. The computational and dynamical properties of the network make it a viable candidate to test the link between Bayesian computations and several physiological features observed in ASD such as inhibitory dysfunction, as well as differences in neural variability and oscillations.

In what follows we will firstly set the basis for this work by recapitulating some of the key findings of Ref.~\citenum{echeveste2020cortical}, relating probabilistic inference, and dynamics in a network model which we will take to describe healthy control subjects. We will then make use of the connection between perception and physiology established by this model and take two parallel routes to explore two different theories for autism: a perceptual theory expressed in terms of hypopriors, and a physiological theory concerning impaired inhibition. The fist path will involve modifying the probabilistic model under which perception takes place, and more concretely its prior, and observing the consequences of that choice in terms of the observer's posteriors. The second path will involve inducing an inhibitory deficit in the neural network whose job is to sample from the corresponding posteriors, and analyzing the effect of that modification in the posteriors represented by the network. We will then compare the results of both approaches to determine to what extent these two seemingly unrelated theories are compatible. Finally, we show that the induced inhibitory deficit in the network model produces changes in the variability and dynamics of the network. We will evaluate these changes in the context of empirical observations in ASD subjects and other theoretical accounts for ASD. These include an increase in neural variability, as well as an increase in the power and frequency of gamma oscillations. The network also becomes hypersensitive to intense stimuli, displaying stronger transients responses at stimulus onset. 

\begin{figure}[h] 
\centering\includegraphics[width=0.6\textwidth]{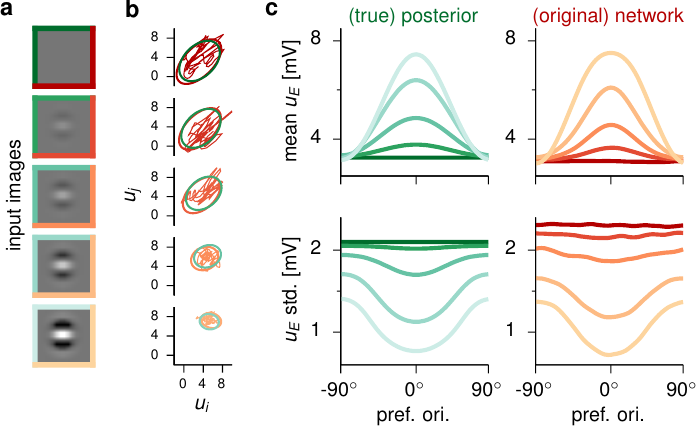}
\caption{\label{fig_original_net}
\textbf{Inference under the GSM and responses in the original network, here representing healthy neurotypical subjects.}
Replotted from Ref.~\citenum{echeveste2020cortical}. In all panels shades of green correspond to the ideal observer, while red corresponds to network responses, as in \Cref{fig_sketch}. Line colors in \textbf{b} and frame colors in \textbf{d} indicate different contrast levels, which are the same as stimulus frames in \textbf{a}, indicating to which stimulus responses correspond. \textbf{a,}~Stimuli (shade of frame color indicates contrast level, split green, blue and red indicates that the same stimuli were used as input to the ideal observer and to both neural networks).  \textbf{b,}~Covariance ellipses ($2$ standard deviations) of the ideal observer's posterior distributions (green) and of the networks' corresponding response distributions (red). Red trajectories show sample $500$~ms-sequences of activities in the networks. As in the sketch of fig.~\Cref{fig_sketch}, 2D projections corresponding to two representative latent variables / excitatory cells are shown. These two correspond to projective fields / receptive fields at preferred orientations $42^\circ$ and $16^\circ$.
\textbf{c,}~Mean (top) and standard deviation (bottom) of latent variable intensities ordered by each latent's orientation, for each stimulus in the training set. Left: from the ideal observer's posterior distribution (green). Right: E cell membrane potentials $u_\e$ from the networks' stationary distributions (red). 
\textbf{d,}~Comparison of correlation matrices. Left: for the ideal observer's posterior distributions (in green). Right: for the networks' stationary response distributions (red). Response moments in \textbf{c} and \textbf{d} were estimated from n = $20,000$ independent samples (taken $200$~ms apart). Correlations in \textbf{d} are Pearson's correlations.
}
\end{figure}

\clearpage

\section*{Results}\label{sec:results}

\subsection*{Bayesian inference of visual features implemented by a recurrent E-I neural circuit}\label{subsec:results_prev}

The starting point for perceptual inference within the Bayesian framework is a probabilistic model that describes one's assumptions about how observed stimuli relate to variables of interest in the outside world. This forward model is usually referred to as a \emph{generative model}, and the role of an ideal Bayesian observer is to invert this probabilistic relationship to obtain posterior distributions over those variables of interest given the observed stimulus. The generative model employed here is a Gaussian Scale-Mixture model (GSM, see \Cref{fig_sketch}~a and \methods), which has been shown to capture the statistics of natural images at the level of small image patches\cite{wainwright2000scale}. Importantly, inference under this model had already been shown to explain features of behavior and stationary response distributions in neural data in visual perception\cite{schwartz2009perceptual,coen2015flexible,orban2016neural}. Under this version of the GSM, natural image patches are constructed as linear combinations of Gabor filters of different orientations, which are then scaled by a global contrast variable. The goal of the inference process was to estimate the probability distribution of the intensity with which each Gabor filter (each orientation) participated in the observed image. In turn, in order to model cortical neural dynamics, a common recurrent neural network model is employed: the stabilized supralinear network (SSN, see \Cref{fig_sketch}~b and \methods)\cite{ahmadian2013analysis,hennequin2018dynamical}. Neurons in the network were arranged around a ring, according to their preferred orientation, under the approximation of visual inference problem being rotationally symmetric (though see \discussion). Moreover, neurons in the network respected Dale's principle, with two separate populations for excitatory (E) and inhibitory (I) cells. The SSN thus formulated was then optimized using current machine learning methods to approximate a Bayesian ideal observer under the GSM: when the network receives an image patch as its input, it produces samples over time with its neural activity so as to represent the corresponding posterior distribution (\Cref{fig_sketch}~c--d). Examples of the image patches used to train the network, as well as sample neural trajectories are presented in \Cref{fig_original_net}~a--b, respectively. After training, posterior distributions sampled by network responses match those prescribed by the ideal observer (see \Cref{fig_original_net}~c, cf. green and red). 
Once trained, the SSN model thus establishes a mechanistic link between neural dynamics in terms of an E-I circuit and perception formulated as sampling-based probabilistic inference. In what follows we exploit this link to take two complementary paths: inducing simple perturbations to the GSM to induce hypopriors, and to the SSN to induce an inhibitory dysfunction.

\subsection*{Perturbing the generative model: the effect of hypopriors}\label{subsec:results_hypo}

\begin{figure}[b!] 
\centering\includegraphics[width=0.6\textwidth]{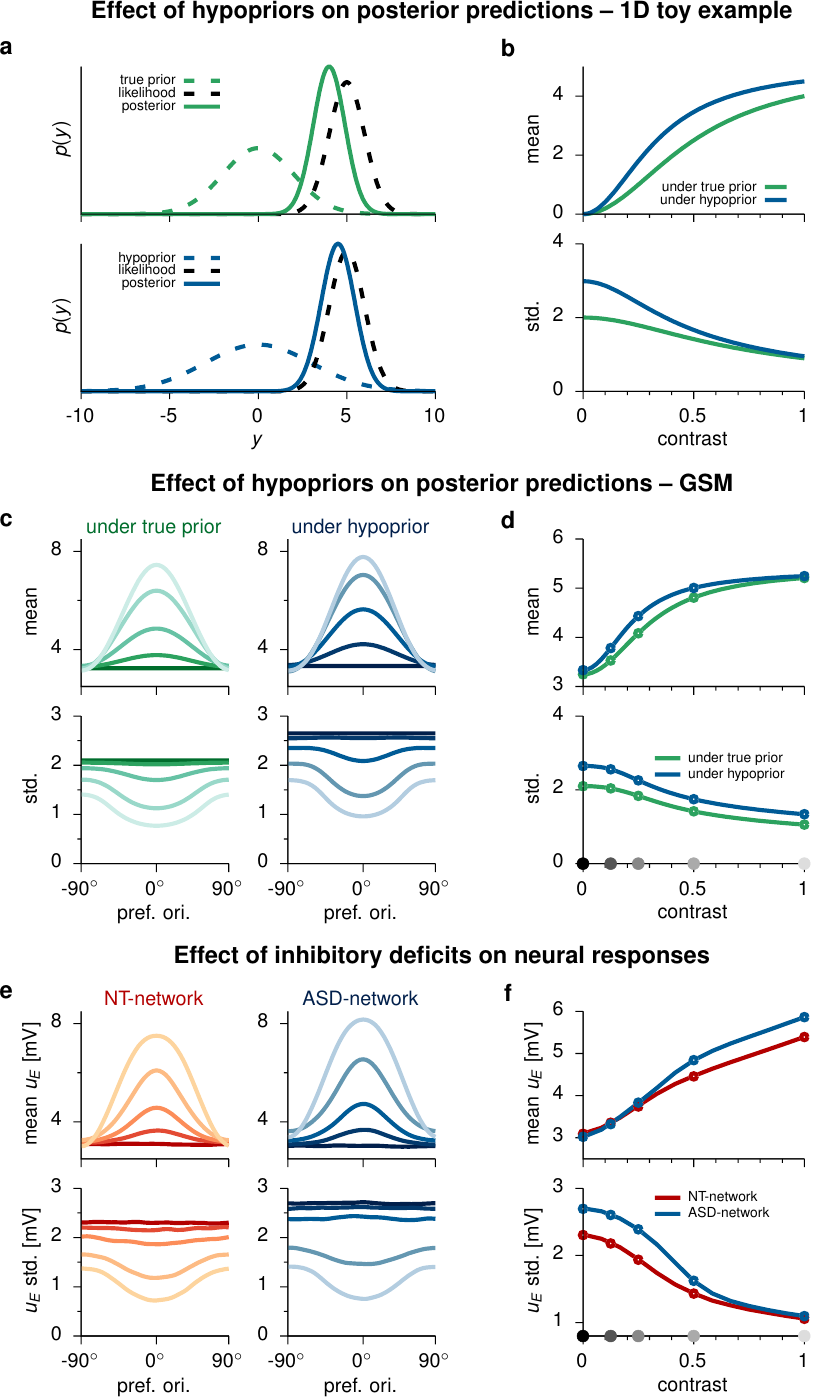}
\caption{\label{fig_main}\textbf{Hypopriors and impaired inhibition.} (Continues on next page)
}
\end{figure}

To illustrate and generate intuitions on the effect of hypopriors, we begin by employing a simplified one-dimensional toy example (\methods). Let us assume the ``true'' prior, correctly describing the statistics of the world concerning a particular inference process, is a zero-mean Gaussian. Let us further assume for this toy example that the likelihood is also a Gaussian function whose precision is modulated by a contrast variable which expresses the degree of reliability of the sensory stimulus. If we vary the stimulus contrast we can compute a posterior distribution for each stimulus under this true prior (\Cref{fig_main}~a~--~b, in green). If, however, we were to employ a hypoprior, that is a prior with a higher variance, we would obtain posterior distributions which overweight sensory evidence, in the sense that they more closely resemble the likelihood function (both in mean and variance) than they should. This in turn results in a higher posterior mean and in higher uncertainty about the estimate (\Cref{fig_main}~b, cf. green and blue lines).

Let us now turn to the GSM. Also in this case, a global contrast variable regulates the reliability of the stimulus. However, in contrast to the 1D toy example presented before, inference in this case takes place in a higher dimensional space. We again modify the prior distribution to induce a hypoprior. We do so in the simplest possible way, by scaling the prior co-variance matrix by a constant factor larger than $1.0$ (\methods). In \Cref{fig_main}~c we compare the posterior distributions calculated under the true prior (in green) with those computed under the hypoprior (in blue). As expected, we again find that hypopriors result in overweighting of sensory stimuli, with higher posterior means and higher uncertainty about the estimates (\Cref{fig_main}~d, cf. green and blue lines), consistently with the postulates of \cite{pellicano2012world}.

\addtocounter{figure}{-1}
\begin{figure} [t!]
  \caption{\textbf{Hypopriors and impaired inhibition.}
\textbf{a--b}~: Effect of hypopriors on posterior predictions for a 1D toy example. 
Priors, likelihoods and posteriors are all Gaussian. A contrast variable regulating the likelihood precision plays the role of the perceptual reliability of stimuli. Two example inference cases are presented: under the true (well-calibrated) prior (dashed, green) and under a wider hypoprior (dashed, blue). \textbf{a}~The prior (dashed, color) and likelihood (dashed, black) are multiplicatively combined according to Bayes' rule to form the posterior (continuous, color). \textbf{b}~Posterior mean (top plot) and standard deviation (bottom plot) under the true prior (green) and the hypoprior (blue), as a function of contrast (likelihood precision).
\textbf{c--d}~: Effect of hypopriors on posterior predictions for the full multivariate GSM model. 
\textbf{c}~Mean (top plots) and standard deviation (bottom plots) of latent variable intensities ordered by each latent's orientation, for each stimulus in \Cref{fig_original_net}. Left: for the well calibrated ideal observer's posterior distribution (green). Right: under a hypoprior (blue).  \textbf{d}~ Posterior mean (Top) and standard deviation (Bottom), averaged across all latent variables, under the true prior (green) and the hypoprior (blue), as a function of contrast.
\textbf{e--f}~: Effect of impaired inhibition on network responses.
\textbf{e,}~Mean (top) and standard deviation (bottom) of latent variable intensities ordered by each latent's orientation, for each stimulus in the training set. E cell membrane potentials $u_\e$ from the stationary response distributions for the NT-network (Left, red), and for the ASD-network (Right, blue). \textbf{f}~ Mean (Top) and standard deviation (Bottom) of neural responses, averaged across all cells, for the NT-network (red) and the ASD-network (blue), as a function of contrast. Circles, and gray dots on x-axis of panels \textbf{d} and \textbf{f} indicate training contrast levels.}%missing
\end{figure}

\subsection*{Perturbing the network: the effect of inhibitory deficits}\label{subsec:results_inh}

We now turn our attention to the network model. In what follows we will refer to the original SSN  presented in \Cref{fig_original_net}, as the \emph{neurotypical} (NT) network. As previously stated, the NT-network was constructed in terms of separate excitatory and inhibitory populations. Here we target inhibitory connections by scaling down their efficacy by a global constant value (\methods). In order to ensure that baseline activity levels are not affected, and following the ideas of Ref.~\citenum{nelson2015excitatory}, we also scaled excitatory connections globally in a homeostatic fashion (see \Cref{fig_params}  and \methods). We will henceforth refer to the network where inhibitory deficits have been induced as the ASD-network. As we did for the generative model, we then compared the mean and standard deviation of the posterior distributions encoded by both networks in terms of their response samples (\Cref{fig_main}~e~--~f). Notably, we observed that ASD-network representations of the posteriors also seemed to overweight current sensory information. Indeed, posterior means were higher in the ASD- than in the NT-network (\Cref{fig_main}~f top panel, cf. red and blue lines). In passing, we note that because of the original approximate inference scheme, the scaling of the mean and standard deviation with contrast between the original network and the posterior are similar but not identical. In particular, while mean responses in the generative model saturate at high contrasts, they only decelerate in the network model, without actually saturating. Indeed, responses in this type of network models do not saturate. They either continue to grow or `bounce back' and begin to decrease\cite{ahmadian2013analysis}. Similarly, a slightly higher standard deviation is observed in the network with respect to the posterior at low contrast, which stems from an underestimation of the variance of neural responses under the Gaussian approximation during training of the network\cite{echeveste2020cortical}.

Higher uncertainty about the estimates was also found in the network (\Cref{fig_main}~f bottom panel, cf. red and blue lines), just as it happened for the generative model under hypopriors (compare \Cref{fig_main} panels d and f). Interestingly, we have reached the same qualitative traits by two very different approaches and following two theories expressed at widely different levels: one perceptual, one physiological.

It is important to note that sampling-based implementations of Bayesian inference establish a direct link between uncertainty and neural variability, since the width of the posterior distribution is directly related to the amount of variability. Indeed we observe that weaker inhibition leads to higher variability in the neural responses of the ASD-network compared to the NT-network (\Cref{fig_main}~f,bottom panel, cf. red and blue lines), as had been suggested in Ref.~\citenum{rubenstein2003model}, where the point had been made that a disruption of E-I balance leading to a hyperexcitable cortex would lead to increased cortical `noise`. Indeed, higher neural variability has been experimentally reported in ASD subjects both in EEG\cite{milne2011increased} and fMRI\cite{haigh2015cortical} studies.

An advantage of employing a neural network model such as the SSN, which shows characteristic features of cortical dynamics, such as gamma oscillations and transient overshoots (including their contrast dependence), is that we can also explore the predictions the model makes for these features, now for the ASD-network. 

\begin{figure}[h] 
\centering\includegraphics[width=\textwidth]{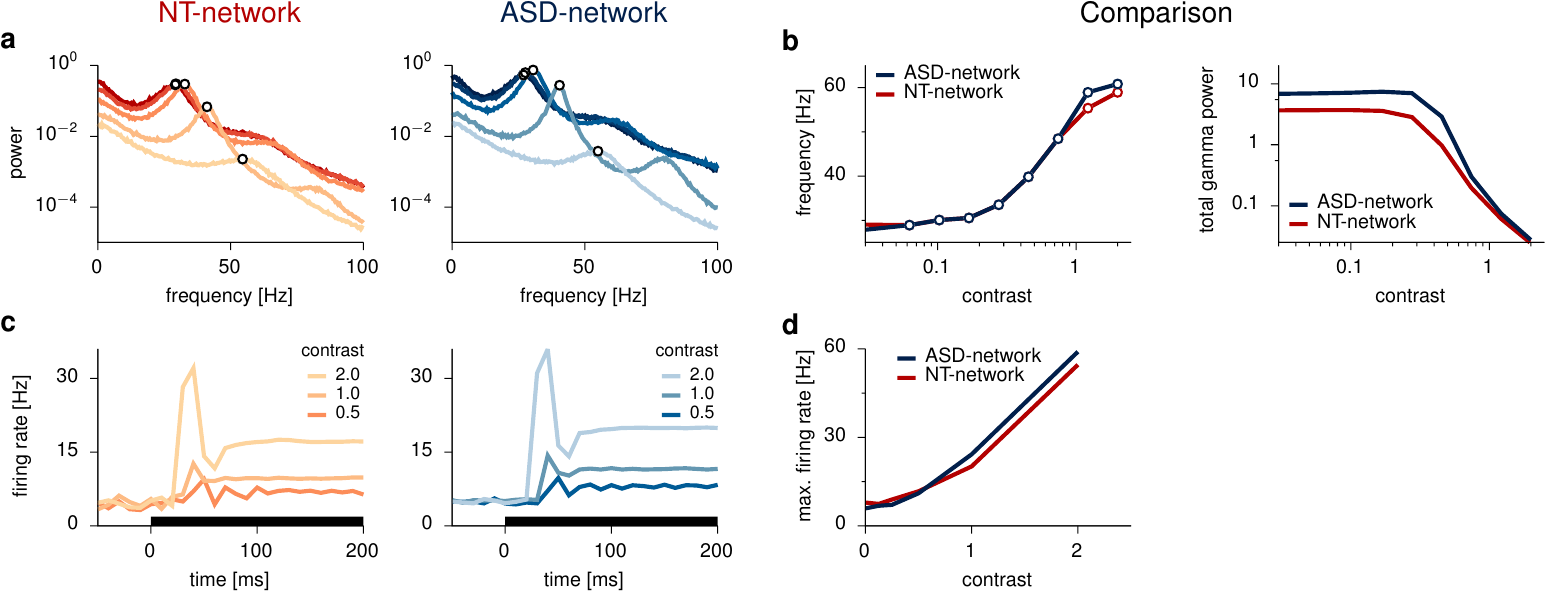}
\caption{\label{fig_transients_and_oscillations}
\textbf{Transient responses and oscillations.}
\textbf{a,}~ LFP power as a function of frequency for stimuli of different contrast levels (same stimuli and colors as in \Cref{fig_main}) in the NT-network (left), and in the ASD-network (right). Both networks present strong gamma oscillations (see peaks in the gamma band, indicated by empty circles). \textbf{b,}~ Comparison of oscillatory behavior in both networks. On the left, the peak gamma frequency is presented as a function of stimulus contrast for both networks. Very minimal differences are observed. On the right, the total power within the gamma band is presented as a function of contrast for both networks. A higher gamma power is observed for the ASD network at all contrasts, with strong differences at low contrasts. \textbf{c,}~ Across-trial average transient responses for stimuli of different contrast levels in the neurotypical network (left) and in the ASD network (righ). Both networks present strong stimulus dependent transient overshoots. \textbf{d,}~ Comparison of overshoot sizes. The maximal firing rate is presented as a function of stimulus contrast for both networks. We observe that the ASD network presents stronger peak responses at higher contrasts, over-reacting to intense stimuli. NT-network results reproduced from Ref.~\citenum{echeveste2020cortical}.
}
\end{figure}

Firstly, we look at gamma oscillations. To that end we computed the power spectrum from the local field potential (LFP), from which we extracted the peak gamma frequency for different contrast levels for both networks (\Cref{fig_transients_and_oscillations}~a). We note that the overall frequency modulation is very similar in both networks, with slightly higher peak gamma frequency in the ASD-network for high contrast stimuli (cf. \Cref{fig_transients_and_oscillations}~b, left panel, red and blue). Previous work has reported higher peak gamma frequency in ASD subjects solving a visual task, which was interpreted as a sign of ``increased neural inhibition''\cite{dickinson2016superior}. At first glance, this might seem at odds with the starting point for our work where we have weakened inhibitory synapses. It is worth noting however that total inputs (both E and I) result in a balanced recurrent network from a dynamic equilibrium, which may result in higher inhibitory currents, despite weaker inhibitory synapses. This is precisely the case here (see \Cref{fig_params} ~d). Indeed, it has been known for decades that balanced networks are prone to so-called ``paradoxical effects''\cite{tsodyks1997paradoxical}, whereby direct external inhibitory inputs to I cells, can actually lead to increased I rates. This also hints at why seemingly contradictory results are often found regarding inhibition in ASD depending on what exactly is chosen as a measure of inhibition.

Interestingly however, gamma power is higher for the ASD-network (see sharper gamma peaks in the spectra of \Cref{fig_transients_and_oscillations}~a, and in \Cref{fig_transients_and_oscillations}~b, right plot, blue vs red). An insight into the functional interpretation of this effect can be obtained from analyzing neural responses at zero contrast, representing what is usually termed spontaneous activity in the literature. In sampling based models, such as this one, spontaneous activity is postulated to encode this prior distribution\cite{berkes2011spontaneous}. Indeed, when the stimulus is completely uninformative, as is the case at zero contrast, the posterior matches the prior. The model hence predicts higher gamma power in spontaneous activity, which is in line with previous reports of higher gamma band power in resting state activity of ASD subjects\cite{van2015increased}.

We finally turn our attention to transient responses. We compared the ASD- and NT-networks in terms of their trial-averaged firing rates around stimulus onset (\Cref{fig_transients_and_oscillations}~c). The model predicts higher maximal firing rates (and not only mean rates) for the ASD network than for the NT network at intermediate and high contrasts (cf. \Cref{fig_transients_and_oscillations}~d, red and blue), indicating that the ASD-network has become hypersensitive to intense stimuli. We note that theories of perception expressed in terms of predictive coding usually interpret peak rates as a measure of surprise, novelty or unexpectedness\cite{rao1999predictive}, and indeed a predictive coding account of ASD perceptual traits, including abnormal sensory sensitivity, has been postulated by several authors in the past\cite{van2013predictive,van2014precise}. Results from the ASD network, which we here interpret from a Bayesian inference perspective, are then not inconsistent with a predictive coding view of perceptual differences in the ASD population.

\section*{Discussion}\label{sec:discussion}

Neural neural network models are increasingly being used as a tool to study how differences in neural architectures may be linked to symptoms in different disorders\cite{lanillos2020review}. In this work we have employed a neural network model of a V1 cortical hypercolumn trained to perform sampling-based probabilistic inference in a visual task to build a mechanistic bridge between descriptions of ASD formulated at two very different levels: a physiological level (in terms of inhibitory dysfunction\cite{rubenstein2003model}, neural variability\cite{milne2011increased,haigh2015cortical}, and gamma oscillation\cite{van2015increased}), and a perceptual level (in terms of hypopriors in Bayesian computations\cite{pellicano2012world}). In what follows we describe merits of this work, limitations and open questions.

\subsection*{Merits}

We have taken two parallel paths: in one perturbing the probabilistic generative model in order to induce hypopriors, and in the other perturbing the neural network model to induce an inhibitory dysfunction. We observed that both approaches lead to consistent results in terms of the represented posterior distributions, providing support for the possibility that both views of ASD might actually constitute two sides of the same coin.

Employing a neural network model such as the SSN, which not only performs inference in a perceptual task but also displays characteristic features of cortical dynamics while doing so\cite{echeveste2020cortical}, allowed us to make further connections between characteristic differences in these dynamics and inhibitory dysfunction in ASD subjects. Stimulus-dependent variability modulations in the network, and concretely the direct link between neural variability and uncertainty established by sampling-based implementations of inference, predicted higher variability in neural responses in the ASD- vs the NT-network. Indeed increased neural variability has been reported in ASD subjects both in EEG\cite{milne2011increased} and fMRI\cite{haigh2015cortical} studies. Moreover, transient overshoots, usually interpreted in predictive coding theories to represent novelty, surprise or unexpectedness\cite{rao1999predictive}, are present in the network, with higher responses for strong stimuli in the ASD-network vs the NT-network, indicating an oversensitivity to intense stimuli, a feature often reported in children with ASD\cite{kern2006pattern}. 

Furthermore, oscillations in the ASD-network displayed higher gamma-band oscillatory power, consistent with observations in resting-state EEG recordings of ASD subjects\cite{van2015increased}. Peak gamma frequencies were also higher in the ASD network for high-contrast stimuli, a fact which has indeed been observed in EEG recordings from subjects performing an orientation discrimination task\cite{dickinson2016superior}, and which had been attributed to increased inhibition. We confirmed that, despite having decreased the efficacy of inhibitory synapses in our network, mean inhibitory inputs were indeed actually larger for high-contrast stimuli. This observation is in line with the known fact that balanced E-I networks are prone to ``paradoxical effects'' regarding inhibition\cite{tsodyks1997paradoxical}, where average rates result from a dynamic balance of excitation and inhibition, and might explain apparent contradictions between studies reporting increased/decreased inhibition\cite{dickinson2016superior,cellot2014gabaergic}. These results also highlight the importance of neural network simulations to assist in the interpretation of physiological observations regarding the role of inhibition in cortical recordings.

\subsection*{Limitations and open questions}

Training recurrent neural networks with expansive non-linearities beyond mean responses is currently a challenging and computationally expensive task. These networks are prone to instabilities and current optimization for second-order moments requires either a large number of trials, or matrix-matrix operations which scale as $n^3$ in the number of neurons\cite{hennequin2016characterizing}. Indeed, the choice of the simple generative model played a key role in order to make the training problem tractable with currently available optimization techniques, but imposes some limitations. The GSM produces multivariate Gaussian posteriors (which enabled training the network with currently available second-order moment-matching methods), and was further constructed to be rotationally symmetric (which drastically reduced the number of network parameters to be optimized, as well as the required number of training examples). A model constructed in this way, will however not be able to capture features of human behavior in popular tests of visual perception, such as the ``oblique effect'', where neurotypical subjects seem to be more sensitive to cardinal orientations\cite{westheimer1998orientation}, an effect which is reduced in ASD subjects\cite{dickinson2014oblique}. Tackling problems like these in a sampling-based setting will require developing tools to train more flexible networks that can produce richer posterior distributions. It should be noted that these limitations are however of a technical nature, and are not inherent to the sampling-based inference framework.

Secondly, the model employed to explain simple, low-level perceptual computations was constructed in terms of a single V1 hypercolumn, and is hence only able to capture local dynamical features, such as locally generated gamma oscillations. Hypothetically, the ideas presented here can be extended to the representation of other circular variables beyond orientation of visual stimuli, such as head direction in rodents\cite{skaggs1995}, motor intent in primates\cite{georgeopoulos1993}, physical space in grid cells \cite{mcnaughton2006}, or oculomotor control\cite{seung1998}. In all these examples, highly specialized brain areas receive assorted inputs that carry a noisy, filtered and distributed representation of a circular variable. The recurrent activity of the network constitutes a mechanistic implementation of an inference process, which could be potentially executed through a sampling-based Bayesian inference strategy, as explored here. If that were the case, the strong reliance of ASD subjects on the likelihood could also be broadened beyond the realm of sensory processing. Extensions of these ideas are also conceivable to other one-dimensional, yet aperiodic, domains, such as sound pitch\cite{aronov2017}, navigation speed\cite{kropff2015}, or elapsed time\cite{tsao2018} which, although still fairly narrow in their semantic content, involve some degree of higher-level processing. However, as we progress into still higher cognitive functions, the understanding of how context-dependent modulations of cortical dynamics emerge during complex perceptual tasks will likely require models where multiple circuits interact\cite{simon2016dysfunction}. In this sense, hierarchical or spatially extended versions of the SSN model employed here may provide adequate substrates to study inference of higher level perceptual tasks where longer-range aspects of cortical dynamics, such as gamma synchronization, might emerge.

Thirdly, we have focused on one aspect of probabilistic inference: inferring the state of a set of latent variables under perceptual uncertainty. The study of other aspects of this problem, such as inferring temporal transitions\cite{sinha2014autism}, or causal relationships\cite{noel2021aberrant}, and their link to altered inhibition and neural dynamics, will require the use of different architectures and generative models and constitute worthwhile avenues of future research.

\subsection*{Closing remarks}

We have shown how recurrent neural networks optimized for sampling-based inference are viable candidates to bridge the gap between Bayesian perceptual theories of ASD and their physiological underpinnings in terms of inhibitory dysfunction, neural variability and oscillations. We believe these results highlight the potential for the use of the emerging body of function-optimized neural networks\cite{yamins2014performance,hennequin2014optimal,song2016training,orhan2017efficient,remington2018flexible,echeveste2020cortical} as models to establish mechanistic links between neural activity and computations in the cortex that go beyond the study of neurotypical perception.

\section*{Methods}\label{sec:methods}

In order to link cortical dynamics and probabilistic computations we modified the parameters of the probabilistic and network models employed in Ref.~\citenum{echeveste2020cortical}. In what follows we describe those changes and refer the reader to the original paper for a more detailed description of the models and of the original model parameters.

\subsection*{The generative model}

In this work the Gaussian scale mixture model (GSM)\cite{wainwright2000scale}, is employed as the generative model of natural images (at the level of small patches) under which inference is carried out in the primary visual cortex (V1)\cite{coen2015flexible,orban2016neural}. Under the GSM an image patch $\xx$ is obtained by linearly combining a number of local features (given by the columns of a matrix $\AA$), which are weighted by a corresponding number of feature coefficients given by $\yy$, further scaled by a single contrast variable $z$, and finally corrupted by additive white Gaussian noise. This forward generative model can then be summarized in terms of the likelihood function given by
\begin{linenomath}
\begin{align}
\xx \cgiven \yy, z &\sim \normal{z \,  \AA \, \yy, \sx \, \II}, \label{eq:pred}
\end{align}
\end{linenomath}
together with the priors for the feature coefficients and the contrast variable $z$. Local features were assumed to be drawn from a multivariate Gaussian:
\begin{linenomath}
\begin{align}
\yy & \sim \normal{\OO, \CC},\hphantom{x} \label{eq:prior}
\end{align}
\end{linenomath}
and the contrast was assumed to be drawn from a Gamma prior. To induce hypopriors we modified the overall scale of the prior covariance matrix $\CC$, by taking $\CC_{\text{HP}} = \alpha_{\text{HP}}\CC$, with $\alpha_{\text{HP}}~=~1.5$. Other values were explored without qualitative differences (not shown). We note that taking $\alpha_{\text{HP}} > 1$ results in wider priors, as required for a hypoprior.

The 1D toy example model of \Cref{fig_main}a--b, corresponds to a 1-dimensional GSM with prior variance $C = 4$, $A = 10$, and $\sx = 100$. As in the full GSM, we took $\alpha_{\text{HP}} = 1.5$.

\subsection*{Network dynamics and architecture}

The circuit model consisted of a nonlinear, stochastic network respecting Dale's principle, with $N_\e$ excitatory and $N_\i$ inhibitory neurons. The evolution of the membrane potential $u_i$ of each neuron $i$ in this model is described by\cite{hennequin2018dynamical}
\begin{linenomath}
\begin{align}
\tau_i \frac{\dd{u_i}}{\dd{t}} &= -\fun{u_i}{t} + \fun{h_i}{t} + {\textstyle \sum_j W_{ij}} \, \fun{r_j}{t} + \fun{\eta_i}{t}, \label{eq:net}
\end{align}
\end{linenomath}
where $\tau_i$ represents the membrane time constant for neuron $i$, $h_i$ its feedforward input, and $\eta_i$ is the \inputnoise{} (capturing both intrinsic and extrinsic forms of neural variability). $\WW$ is the matrix of recurrent connections, and hence $W_{ij}$ represents the  strength of the synapse connecting neuron $j$ to neuron $i$. As previously mentioned, the network is non-linear, with firing rates
\begin{linenomath}
\begin{align}
\fun{r_i}{t} &= k \floor*{\fun{u_i}{t}}^m. \label{eq:nonlin}
\end{align}
\end{linenomath}
Here $k$ and $m$ represent the scale and exponent of the firing rate nonlinearity\cite{ahmadian2013analysis}. Given the rotational symmetry of the problem, $\WW$ itself was parametrized to be rotationally symmetric. Neurons in the model are arranged in a ring of pairs of E and I cells according to their preferred orientations (\Cref{fig_sketch}c) where $W_{ij}$ was a smoothly decaying function of the tuning difference between neurons $i$ and $j$ (see \Cref{fig_params} ~a, top and second row). The (stimulus-independent) \inputnoise{} covariance was analogously parametrized (see \Cref{fig_params} ~a, third row). Following canonical models of V1 simple cells\cite{dayan2001theoretical}, feedforward inputs to the network were computed by applying a linear filter $\WW^{\text{ff}}$ to the stimulus (the image patch) followed by a nonlinearity (see \Cref{fig_params} ~a, bottom row).

The perturbation here employed to induce an inhibitory deficit has a single free parameter $\delta_\text{I}$ which scales the inhibitory columns of $\WW$, $\WW_\text{I}^\text{ASD} = (1 - \delta_\text{I}) \WW_\text{I}^\text{NT}$ (see \Cref{fig_params} ~a--b). In order to maintain the baseline level of activity, a second modification is introduced (simulating homeostatic adaption of the excitatory connections), scaling the excitatory columns of $\WW$ by a factor $\delta_\text{E}$: $\WW_\text{E}^\text{ASD} = (1 - \delta_\text{E}) \WW_\text{E}^\text{NT}$. This second factor was found by grid-search minimization of the homeostatic cost
\begin{linenomath}
\begin{align}
\mathcal{C}_\text{h} &= \abs{\mu_{\text{s}}^{\text{NT}}-\mu_{\text{s}}^{\text{ASD}}}, \label{eq:homeost}
\end{align}
\end{linenomath}
capturing the change in mean spontaneous activity levels ($\mu_{\text{s}}$) between the original NT- and perturbed ASD-network . This adaptation procedure returns a single $\delta_\text{E}$ value for each $\delta_\text{I}$ value (\Cref{fig_params} ~c). We note that excitatory changes via this procedure resulted always smaller than inhibitory ones (cf. to identity line in \Cref{fig_params} ~c, bottom plot). Network results presented throughout this paper correspond to $\delta_\text{I} = 0.1$, for which $\delta_\text{E} = 0.076$. Numerical experiments were repeated for $\delta_\text{I} = 0.05$ and $\delta_\text{I} = 0.15$ without qualitative differences (not shown).

\subsection*{Numerical simulations}

Stationary moments of neural responses to a fixed input (\Cref{fig_main}e) were computed from $20,000$ independent samples ($200$~ms apart) generated by letting neural activity in the network evolve over time via \Cref{eq:net} (excluding transients).  Power spectra in \Cref{fig_transients_and_oscillations}~a were obtained from simulated local field potentials (LFPs), computed as the average (across-cells) membrane potential. Gamma peak frequencies in \Cref{fig_transients_and_oscillations}~b (left) were obtained as the local maximum in the spectrum within the gamma range ($20$--$80$~Hz), while total gamma power in \Cref{fig_transients_and_oscillations}~b (right) was computed as the integral of the spectrum over that same range. 

Transient responses displayed in \Cref{fig_transients_and_oscillations}~c were computed as the mean (across E-cells and trials) firing rates ($n=100$), which are then further averaged over a $10$-ms sliding window. A random delay time (sampled from a truncated Gaussian, with a mean of $45$~ms and a standard deviation of $5$~ms) was employed for the feedforward input to each pair of E--I cells. These procedures had been put in place to allow for a comparison to experimental data, and are here kept in order to compare the ASD-netowork to replotted results from the original (here NT-) network. Maximal firing rates in \Cref{fig_transients_and_oscillations}~d were obtained as the peak rates from transient firing rate responses.

\subsection*{Code availability.}

The (Python) code to create the ASD network is provided in 
\href{https://bitbucket.org/RSE_1987/inhibitory_dysfunction}{\nolinkurl{bitbucket.org/RSE_1987/inhibitory_dysfunction}}.
The code for the numerical experiments can be found at:
\href{https://bitbucket.org/RSE_1987/ssn_inference_numerical_experiments}{\nolinkurl{bitbucket.org/RSE_1987/ssn_inference_numerical_experiments}}.

\subsection*{Acknowledgements}

This work was supported by Argentina's National Scientific and Technical Research Council (\mbox{CONICET}), who covered all researchers salaries. We are grateful to Y. Nagai for pointing out this potential avenue of research after discussing previous work.

\bibliographystyle{unsrt}
\bibliography{main}

\begin{thebibliography}{10}

\bibitem{american2013diagnostic}
American~Psychiatric Association.
\newblock {\em Diagnostic and statistical manual of mental disorders
  (DSM-5{\textregistered})}.
\newblock American Psychiatric Pub, 2013.

\bibitem{ozonoff2008onset}
Sally Ozonoff, Kelly Heung, Robert Byrd, Robin Hansen, and Irva
  Hertz-Picciotto.
\newblock The onset of autism: patterns of symptom emergence in the first years
  of life.
\newblock {\em Autism research}, 1(6):320--328, 2008.

\bibitem{rubenstein2003model}
JLR Rubenstein and Michael~M Merzenich.
\newblock Model of autism: increased ratio of excitation/inhibition in key
  neural systems.
\newblock {\em Genes, Brain and Behavior}, 2(5):255--267, 2003.

\bibitem{cellot2014gabaergic}
Giada Cellot and Enrico Cherubini.
\newblock Gabaergic signaling as therapeutic target for autism spectrum
  disorders.
\newblock {\em Frontiers in pediatrics}, 2:70, 2014.

\bibitem{bolton2011epilepsy}
Patrick~F Bolton, Iris Carcani-Rathwell, Jane Hutton, Sue Goode, Patricia
  Howlin, and Michael Rutter.
\newblock Epilepsy in autism: features and correlates.
\newblock {\em The British Journal of Psychiatry}, 198(4):289--294, 2011.

\bibitem{robertson2016reduced}
Caroline~E Robertson, Eva-Maria Ratai, and Nancy Kanwisher.
\newblock Reduced gabaergic action in the autistic brain.
\newblock {\em Current Biology}, 26(1):80--85, 2016.

\bibitem{horder2018gabaa}
Jamie Horder, Max Andersson, Maria~A. Mendez, Nisha Singh, \"{A}mma Tangen,
  Johan Lundberg, Antony Gee, Christer Halldin, Mattia Veronese, Sven
  B\"{o}lte, Lars Farde, Teresa Sementa, Diana Cash, Karen Higgins, Debbie
  Spain, Federico Turkheimer, Inge Mick, Sudhakar Selvaraj, David~J. Nutt, Anne
  Lingford-Hughes, Oliver~D. Howes, Declan~G. Murphy, and Jacqueline Borg.
\newblock Gabaa receptor availability is not altered in adults with autism
  spectrum disorder or in mouse models.
\newblock {\em Science Translational Medicine}, 10(461), 2018.

\bibitem{nelson2015excitatory}
Sacha~B Nelson and Vera Valakh.
\newblock Excitatory/inhibitory balance and circuit homeostasis in autism
  spectrum disorders.
\newblock {\em Neuron}, 87(4):684--698, 2015.

\bibitem{happe2006weak}
Francesca Happ{\'e} and Uta Frith.
\newblock The weak coherence account: detail-focused cognitive style in autism
  spectrum disorders.
\newblock {\em Journal of autism and developmental disorders}, 36(1):5--25,
  2006.

\bibitem{mottron2006enhanced}
Laurent Mottron, Michelle Dawson, Isabelle Soulieres, Benedicte Hubert, and
  Jake Burack.
\newblock Enhanced perceptual functioning in autism: an update, and eight
  principles of autistic perception.
\newblock {\em Journal of autism and developmental disorders}, 36(1):27--43,
  2006.

\bibitem{palmer2017bayesian}
Colin~J Palmer, Rebecca~P Lawson, and Jakob Hohwy.
\newblock Bayesian approaches to autism: Towards volatility, action, and
  behavior.
\newblock {\em Psychological bulletin}, 143(5):521, 2017.

\bibitem{van2013predictive}
Jeroen~JA Van~Boxtel and Hongjing Lu.
\newblock A predictive coding perspective on autism spectrum disorders.
\newblock {\em Frontiers in psychology}, 4:19, 2013.

\bibitem{van2014precise}
Sander Van~de Cruys, Kris Evers, Ruth Van~der Hallen, Lien Van~Eylen, Bart
  Boets, Lee de~Wit, and Johan Wagemans.
\newblock Precise minds in uncertain worlds: predictive coding in autism.
\newblock {\em Psychological review}, 121(4):649, 2014.

\bibitem{knill1996perception}
DC~Knill and W~Richards.
\newblock {\em Perception as {Bayesian} inference}.
\newblock Cambridge University Press, 1996.

\bibitem{fiser2010statistically}
J~Fiser, P~Berkes, G~Orb{\'a}n, and M~Lengyel.
\newblock Statistically optimal perception and learning: from behavior to
  neural representations.
\newblock {\em Trends in {Cognitive} {Sciences}}, 14(3):119--130, 2010.

\bibitem{pellicano2012world}
Elizabeth Pellicano and David Burr.
\newblock When the world becomes ‘too real’: a bayesian explanation of
  autistic perception.
\newblock {\em Trends in cognitive sciences}, 16(10):504--510, 2012.

\bibitem{rao1999predictive}
RPN Rao and DH~Ballard.
\newblock Predictive coding in the visual cortex: a functional interpretation
  of some extra-classical receptive-field effects.
\newblock {\em Nature Neuroscience}, 2(1):79, 1999.

\bibitem{aitchison2017or}
L~Aitchison and M~Lengyel.
\newblock With or without you: predictive coding and bayesian inference in the
  brain.
\newblock {\em Current opinion in neurobiology}, 46:219--227, 2017.

\bibitem{aitchison2016hamiltonian}
L~Aitchison and M~Lengyel.
\newblock The {Hamiltonian} brain: efficient probabilistic inference with
  excitatory-inhibitory neural circuit dynamics.
\newblock {\em PLoS computational biology}, 12(12):e1005186, 2016.

\bibitem{echeveste2020cortical}
Rodrigo Echeveste, Laurence Aitchison, Guillaume Hennequin, and M{\'a}t{\'e}
  Lengyel.
\newblock Cortical-like dynamics in recurrent circuits optimized for
  sampling-based probabilistic inference.
\newblock {\em Nature Neuroscience}, 23(9):1138--1149, 2020.

\bibitem{bastos2012canonical}
AM~Bastos, WM~Usrey, RA~Adams, GR~Mangun, P~Fries, and KJ~Friston.
\newblock Canonical microcircuits for predictive coding.
\newblock {\em Neuron}, 76(4):695--711, 2012.

\bibitem{roberts2013robust}
MJ~Roberts, E~Lowet, NM~Brunet, M~Ter~Wal, P~Tiesinga, P~Fries, and P~De~Weerd.
\newblock Robust gamma coherence between macaque {V1} and {V2} by dynamic
  frequency matching.
\newblock {\em Neuron}, 78(3):523--536, 2013.

\bibitem{churchland2010stimulus}
Mark~M Churchland, Byron~M Yu, John~P Cunningham, Leo~P Sugrue, Marlene~R
  Cohen, Greg~S Corrado, William~T Newsome, Andrew~M Clark, Paymon Hosseini,
  Benjamin~B Scott, David~C Bradley, Matthew~A Smith, Adam Kohn, J~Anthony
  Movshon, Katherine~M Armstrong, Tirin Moore, Steve~W Chang, Lawrence~H
  Snyder, Stephen~G Lisberger, Nicholas~J Priebe, Ian~M Finn, David Ferster,
  Stephen~I Ryu, Gopal Santhanam, Maneesh Sahani, and Krishna~V Shenoy.
\newblock Stimulus onset quenches neural variability: a widespread cortical
  phenomenon.
\newblock {\em Nature Neuroscience}, 13(3):369, 2010.

\bibitem{orban2016neural}
G~Orb{\'a}n, P~Berkes, J~Fiser, and M~Lengyel.
\newblock Neural variability and sampling-based probabilistic representations
  in the visual cortex.
\newblock {\em Neuron}, 92(2):530--543, 2016.

\bibitem{ma2006bayesian}
WJ~Ma, JM~Beck, PE~Latham, and A~Pouget.
\newblock {Bayesian} inference with probabilistic population codes.
\newblock {\em Nature Neuroscience}, 9(11):1432--1438, 2006.

\bibitem{rosenberg2015computational}
Ari Rosenberg, Jaclyn~Sky Patterson, and Dora~E Angelaki.
\newblock A computational perspective on autism.
\newblock {\em Proceedings of the National Academy of Sciences},
  112(30):9158--9165, 2015.

\bibitem{westheimer1998orientation}
Gerald Westheimer and Bettina~L Beard.
\newblock Orientation dependency for foveal line stimuli: detection and
  intensity discrimination, resolution, orientation discrimination and vernier
  acuity.
\newblock {\em Vision research}, 38(8):1097--1103, 1998.

\bibitem{dickinson2014oblique}
Abigail Dickinson, Myles Jones, and Elizabeth Milne.
\newblock Oblique orientation discrimination thresholds are superior in those
  with a high level of autistic traits.
\newblock {\em Journal of autism and developmental disorders},
  44(11):2844--2850, 2014.

\bibitem{milne2011increased}
Elizabeth Milne.
\newblock Increased intra-participant variability in children with autistic
  spectrum disorders: evidence from single-trial analysis of evoked eeg.
\newblock {\em Frontiers in psychology}, 2:51, 2011.

\bibitem{haigh2015cortical}
Sarah~M Haigh, David~J Heeger, Ilan Dinstein, Nancy Minshew, and Marlene
  Behrmann.
\newblock Cortical variability in the sensory-evoked response in autism.
\newblock {\em Journal of autism and developmental disorders},
  45(5):1176--1190, 2015.

\bibitem{van2015increased}
Eric van Diessen, Joeky Senders, Floor~E Jansen, Maria Boersma, and Hilgo
  Bruining.
\newblock Increased power of resting-state gamma oscillations in autism
  spectrum disorder detected by routine electroencephalography.
\newblock {\em European archives of psychiatry and clinical neuroscience},
  265(6):537--540, 2015.

\bibitem{berkes2011spontaneous}
P~Berkes, G~Orb{\'a}n, M~Lengyel, and J~Fiser.
\newblock Spontaneous cortical activity reveals hallmarks of an optimal
  internal model of the environment.
\newblock {\em Science}, 331(6013):83--87, 2011.

\bibitem{haefner2016perceptual}
RM~Haefner, P~Berkes, and J~Fiser.
\newblock Perceptual decision-making as probabilistic inference by neural
  sampling.
\newblock {\em Neuron}, 90(3):649--660, 2016.

\bibitem{haider2013inhibition}
B~Haider, M~H{\"a}usser, and M~Carandini.
\newblock Inhibition dominates sensory responses in the awake cortex.
\newblock {\em Nature}, 493(7430):97--100, 2013.

\bibitem{ray2010differences}
S~Ray and John~HR Maunsell.
\newblock Differences in gamma frequencies across visual cortex restrict their
  possible use in computation.
\newblock {\em Neuron}, 67(5):885--896, 2010.

\bibitem{carandini2012normalization}
M~Carandini and DJ~Heeger.
\newblock Normalization as a canonical neural computation.
\newblock {\em Nature Reviews Neuroscience}, 13(1):51, 2012.

\bibitem{henaff2020representation}
Olivier~J H{\'e}naff, Zoe~M Boundy-Singer, Kristof Meding, Corey~M Ziemba, and
  Robbe~LT Goris.
\newblock Representation of visual uncertainty through neural gain variability.
\newblock {\em Nature communications}, 11(1):1--12, 2020.

\bibitem{wainwright2000scale}
MJ~Wainwright and EP~Simoncelli.
\newblock Scale mixtures of {Gaussians} and the statistics of natural images.
\newblock In {\em Advances in {Neural} {Information} {Processing} {Systems}},
  pages 855--861, 2000.

\bibitem{schwartz2009perceptual}
O~Schwartz, TJ~Sejnowski, and P~Dayan.
\newblock Perceptual organization in the tilt illusion.
\newblock {\em Journal of Vision}, 9(4):19--19, 2009.

\bibitem{coen2015flexible}
R~Coen-Cagli, A~Kohn, and O~Schwartz.
\newblock Flexible gating of contextual influences in natural vision.
\newblock {\em Nature Neuroscience}, 2015.

\bibitem{ahmadian2013analysis}
Y~Ahmadian, DB~Rubin, and KD~Miller.
\newblock Analysis of the stabilized supralinear network.
\newblock {\em Neural Computation}, 25(8):1994--2037, 2013.

\bibitem{hennequin2018dynamical}
G~Hennequin, Y~Ahmadian, DB~Rubin, M~Lengyel, and KD~Miller.
\newblock The dynamical regime of sensory cortex: stable dynamics around a
  single stimulus-tuned attractor account for patterns of noise variability.
\newblock {\em Neuron}, 98(4):846--860, 2018.

\bibitem{dickinson2016superior}
Abigail Dickinson, Michael Bruyns-Haylett, Richard Smith, Myles Jones, and
  Elizabeth Milne.
\newblock Superior orientation discrimination and increased peak gamma
  frequency in autism spectrum conditions.
\newblock {\em Journal of abnormal psychology}, 125(3):412, 2016.

\bibitem{tsodyks1997paradoxical}
Misha~V Tsodyks, William~E Skaggs, Terrence~J Sejnowski, and Bruce~L
  McNaughton.
\newblock Paradoxical effects of external modulation of inhibitory
  interneurons.
\newblock {\em Journal of neuroscience}, 17(11):4382--4388, 1997.

\bibitem{lanillos2020review}
Pablo Lanillos, Daniel Oliva, Anja Philippsen, Yuichi Yamashita, Yukie Nagai,
  and Gordon Cheng.
\newblock A review on neural network models of schizophrenia and autism
  spectrum disorder.
\newblock {\em Neural Networks}, 122:338--363, 2020.

\bibitem{kern2006pattern}
Janet~K Kern, Madhukar~H Trivedi, Carolyn~R Garver, Bruce~D Grannemann,
  Alonzo~A Andrews, Jayshree~S Savla, Danny~G Johnson, Jyutika~A Mehta, and
  Jennifer~L Schroeder.
\newblock The pattern of sensory processing abnormalities in autism.
\newblock {\em Autism}, 10(5):480--494, 2006.

\bibitem{hennequin2016characterizing}
G~Hennequin and M~Lengyel.
\newblock Characterizing variability in nonlinear recurrent neuronal networks.
\newblock {\em arXiv preprint arXiv:1610.03110}, 2016.

\bibitem{skaggs1995}
W.~E. Skaggs, J.~J. Knierim, H.~S. Kudrimoti, and B.~L. McNaughton.
\newblock A model of the neural basis of the rat's sense of direction.
\newblock {\em Advances in Neural Information Processing Systems}, 7:173--180,
  1995.

\bibitem{georgeopoulos1993}
A.~P. Georgopoulos, M.~Taira, and A.~Lukashin.
\newblock Cognitive neurophysiology of the motor cortex.
\newblock {\em Science}, 260(5104):47--52, 1993.

\bibitem{mcnaughton2006}
Bruce~L. McNaughton, Francesco~P. Battaglia, Ole Jensen, Edvard~I. Moser, and
  May-Britt Moser.
\newblock Path integration and the neural basis of the ``cognitive map''.
\newblock {\em Nature Reviews in Neuroscience}, 7(8):663--678, 2006.

\bibitem{seung1998}
Sebastian Seung.
\newblock Cognitive neurophysiology of the motor cortex.
\newblock {\em Neural Networks}, 11(7-8):1253--1258, 1998.

\bibitem{aronov2017}
D.~Aronov, R.~Nevers, and D.~W. Tank.
\newblock Mapping of a non-spatial dimension by the hippocampal-entorhinal
  circuit.
\newblock {\em Nature}, 543(7647):719--722, 2017.

\bibitem{kropff2015}
Emilio Kropff, James~E. Carmichael, May-Britt Moser, and Edvard~I. Moser.
\newblock Speed cells in the medial entorhinal cortex.
\newblock {\em Nature}, 523(7561):419--424, 2015.

\bibitem{tsao2018}
A.~Tsao, J.~Sugar, L.~Lu, C.~Wang, J.~J. Knierim, M.~B. Moser, and E.~I. Moser.
\newblock Integrating time from experience in the lateral entorhinal cortex.
\newblock {\em Nature}, 561(7721):57--62, 2018.

\bibitem{simon2016dysfunction}
David~M Simon and Mark~T Wallace.
\newblock Dysfunction of sensory oscillations in autism spectrum disorder.
\newblock {\em Neuroscience \& Biobehavioral Reviews}, 68:848--861, 2016.

\bibitem{sinha2014autism}
Pawan Sinha, Margaret~M Kjelgaard, Tapan~K Gandhi, Kleovoulos Tsourides,
  Annie~L Cardinaux, Dimitrios Pantazis, Sidney~P Diamond, and Richard~M Held.
\newblock Autism as a disorder of prediction.
\newblock {\em Proceedings of the National Academy of Sciences},
  111(42):15220--15225, 2014.

\bibitem{noel2021aberrant}
Jean-Paul Noel, Sabyasachi Shivkumar, Kalpana Dokka, Ralf Haefner, and Dora
  Angelaki.
\newblock Aberrant causal inference and presence of a compensatory mechanism in
  autism spectrum disorder.
\newblock {\em PsyArXiv}, 2021.

\bibitem{yamins2014performance}
DLK Yamins, H~Hong, CF~Cadieu, EA~Solomon, D~Seibert, and JJ~DiCarlo.
\newblock Performance-optimized hierarchical models predict neural responses in
  higher visual cortex.
\newblock {\em Proceedings of the National Academy of Sciences},
  111(23):8619--8624, 2014.

\bibitem{hennequin2014optimal}
G~Hennequin, TP~Vogels, and W~Gerstner.
\newblock Optimal control of transient dynamics in balanced networks supports
  generation of complex movements.
\newblock {\em Neuron}, 82(6):1394--1406, 2014.

\bibitem{song2016training}
HF~Song, GR~Yang, and XJ~Wang.
\newblock Training excitatory-inhibitory recurrent neural networks for
  cognitive tasks: A simple and flexible framework.
\newblock {\em PLoS computational biology}, 12(2):e1004792, 2016.

\bibitem{orhan2017efficient}
AE~Orhan and WJ~Ma.
\newblock Efficient probabilistic inference in generic neural networks trained
  with non-probabilistic feedback.
\newblock {\em Nature communications}, 8(1):138, 2017.

\bibitem{remington2018flexible}
ED~Remington, D~Narain, EA~Hosseini, and M~Jazayeri.
\newblock Flexible sensorimotor computations through rapid reconfiguration of
  cortical dynamics.
\newblock {\em Neuron}, 98(5):1005--1019, 2018.

\bibitem{dayan2001theoretical}
P~Dayan and LF~Abbott.
\newblock {\em Theoretical neuroscience}, volume 806.
\newblock Cambridge, MA: MIT Press, 2001.

\end{thebibliography}

\clearpage

\begin{efigure}[t!]
\centering
\includegraphics[width=0.6\textwidth]{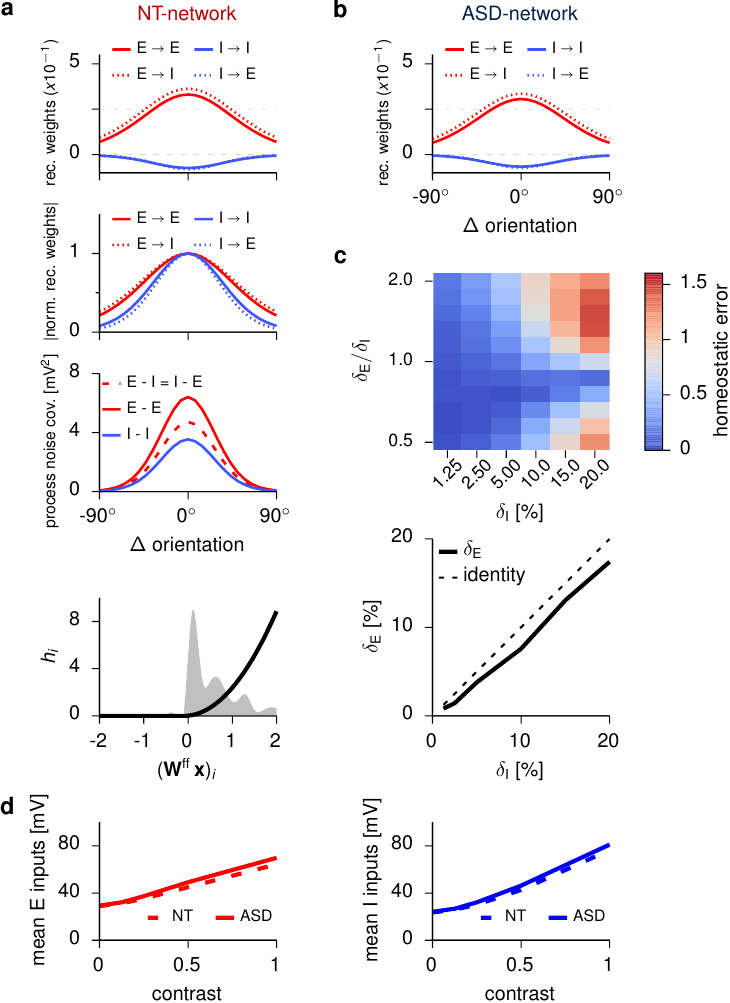}
\caption{\label{fig_params}
\textbf{a,}~ Parameters of the NT-network. Given the symmetry of the problem, both the synaptic weights (top two rows) and the input noise covariance (third row) are given by circulant matrices constructed by the templates presented here. The bottom row depicts the input nonlinearity (in black) and, as a reference, the distribution of inputs across all cells for the training set. \textbf{b,}~ Recurrent weight templates for the perturbed neural network, regarded here as the ASD network. All other parameters are shared with the NT-network \textbf{c,}~ Homeostatic adjustment of excitation. Top: Color plot represents homeostatic error (deviation of baseline activity levels) as a function of the ratio of changes in excitation and inhibition ($\delta_E/\delta_I$), for different choices of $\delta_I$. Bottom: For each value of $\delta_I$, the value of $\delta_E$ which minimizes the homeostatic error in the top plot is selected via grid search. The dashed line corresponds to the identity function. The ASD network presented in \textbf{b}, and employed throughout this work corresponds to the $\delta_I = 10\%$ case. \textbf{d,}~ Average total recurrent excitatory (left) and inhibitory (right) inputs to the ASD-network (full lines) and the NT-network (dashed lines) as a function of contrast.
}
\end{efigure}

\end{document}